\documentstyle[twoside,epsfig]{article}
\catcode`\@=11
\long\def\@makefntext#1{
\protect\noindent \hbox to 3.2pt {\hskip-.9pt
$^{{\eightrm\@thefnmark}}$\hfil}#1\hfill}		%CAN BE USED

\def\@makefnmark{\hbox to 0pt{$^{\@thefnmark}$\hss}}	%ORIGINAL

\def\ps@myheadings{\let\@mkboth\@gobbletwo
\def\@oddhead{\hbox{}
\rightmark\hfil\eightrm\thepage}
\def\@oddfoot{}\def\@evenhead{\eightrm\thepage\hfil
\leftmark\hbox{}}\def\@evenfoot{}
\def\sectionmark##1{}\def\subsectionmark##1{}}

\oddsidemargin=\evensidemargin
\newcounter{sectionc}\newcounter{subsectionc}\newcounter{subsubsectionc}
\renewcommand{\section}[1] {\vspace{12pt}\addtocounter{sectionc}{1}
\setcounter{subsectionc}{0}\setcounter{subsubsectionc}{0}\noindent
	{\tenbf\thesectionc. #1}\par\vspace{5pt}}
\renewcommand{\subsection}[1] {\vspace{12pt}\addtocounter{subsectionc}{1}
	\setcounter{subsubsectionc}{0}\noindent
	{\bf\thesectionc.\thesubsectionc. {\kern1pt \bfit #1}}\par\vspace{5pt}}
\renewcommand{\subsubsection}[1] {\vspace{12pt}\addtocounter{subsubsectionc}{1}
	\noindent{\tenrm\thesectionc.\thesubsectionc.\thesubsubsectionc.
	{\kern1pt \tenit #1}}\par\vspace{5pt}}
\newcommand{\nonumsection}[1] {\vspace{12pt}\noindent{\tenbf #1}
	\par\vspace{5pt}}

\newcounter{appendixc}
\newcounter{subappendixc}[appendixc]
\newcounter{subsubappendixc}[subappendixc]
\renewcommand{\thesubappendixc}{\Alph{appendixc}.\arabic{subappendixc}}
\renewcommand{\thesubsubappendixc}
	{\Alph{appendixc}.\arabic{subappendixc}.\arabic{subsubappendixc}}

\renewcommand{\appendix}[1] {\vspace{12pt}
        \refstepcounter{appendixc}
        \setcounter{figure}{0}
        \setcounter{table}{0}
        \setcounter{lemma}{0}
        \setcounter{theorem}{0}
        \setcounter{corollary}{0}
        \setcounter{definition}{0}
        \setcounter{equation}{0}
        \renewcommand{\thefigure}{\Alph{appendixc}.\arabic{figure}}
        \renewcommand{\thetable}{\Alph{appendixc}.\arabic{table}}
        \renewcommand{\theappendixc}{\Alph{appendixc}}
        \renewcommand{\thelemma}{\Alph{appendixc}.\arabic{lemma}}
        \renewcommand{\thetheorem}{\Alph{appendixc}.\arabic{theorem}}
        \renewcommand{\thedefinition}{\Alph{appendixc}.\arabic{definition}}
        \renewcommand{\thecorollary}{\Alph{appendixc}.\arabic{corollary}}
        \renewcommand{\theequation}{\Alph{appendixc}.\arabic{equation}}
        \noindent{\tenbf Appendix \theappendixc #1}\par\vspace{5pt}}
\newcommand{\subappendix}[1] {\vspace{12pt}
        \refstepcounter{subappendixc}
        \noindent{\bf Appendix \thesubappendixc. {\kern1pt \bfit #1}}
	\par\vspace{5pt}}
\newcommand{\subsubappendix}[1] {\vspace{12pt}
        \refstepcounter{subsubappendixc}
        \noindent{\rm Appendix \thesubsubappendixc. {\kern1pt \tenit #1}}
	\par\vspace{5pt}}

\newcommand{\textlineskip}{\baselineskip=13pt}
\newcommand{\smalllineskip}{\baselineskip=10pt}

\def\eightcirc{
\begin{picture}(0,0)
\put(4.4,1.8){\circle{6.5}}
\end{picture}}
\def\eightcopyright{\eightcirc\kern2.7pt\hbox{\eightrm c}}

\newcommand{\copyrightheading}[1]
	{\vspace*{-2.5cm}\smalllineskip{\flushleft
	{\footnotesize Modern Physics Letters A, #1}\\
	{\footnotesize $\eightcopyright$\, World Scientific Publishing
	 Company}\\
	 }}

\def\abstracts#1#2#3{{
	\centering{\begin{minipage}{4.5in}\footnotesize\baselineskip=10pt
	\parindent=0pt #1\par
	\parindent=15pt #2\par
	\parindent=15pt #3
	\end{minipage}}\par}}

\newcommand{\bibit}{\nineit}
\newcommand{\bibbf}{\ninebf}
\renewenvironment{thebibliography}[1]
	{\frenchspacing
	 \ninerm\baselineskip=11pt
	 \begin{list}{\arabic{enumi}.}
        {\usecounter{enumi}\setlength{\parsep}{0pt}
	 \setlength{\leftmargin 12.7pt}{\rightmargin 0pt} %FOR 1--9 ITEMS
         \setlength{\itemsep}{0pt} \settowidth
	{\labelwidth}{#1.}\sloppy}}{\end{list}}

\newcounter{itemlistc}
\newcounter{romanlistc}
\newcounter{alphlistc}
\newcounter{arabiclistc}

\newcommand{\fcaption}[1]{
        \refstepcounter{figure}
        \setbox\@tempboxa = \hbox{\footnotesize Fig.~\thefigure. #1}
        \ifdim \wd\@tempboxa > 5in
           {\begin{center}
        \parbox{5in}{\footnotesize\smalllineskip Fig.~\thefigure. #1}
            \end{center}}
        \else
             {\begin{center}
             {\footnotesize Fig.~\thefigure. #1}
              \end{center}}
        \fi}

\newcommand{\tcaption}[1]{
        \refstepcounter{table}
        \setbox\@tempboxa = \hbox{\footnotesize Table~\thetable. #1}
        \ifdim \wd\@tempboxa > 5in
           {\begin{center}
        \parbox{5in}{\footnotesize\smalllineskip Table~\thetable. #1}
            \end{center}}
        \else
             {\begin{center}
             {\footnotesize Table~\thetable. #1}
              \end{center}}
        \fi}

\def\@citex[#1]#2{\if@filesw\immediate\write\@auxout
	{\string\citation{#2}}\fi
\def\@citea{}\@cite{\@for\@citeb:=#2\do
	{\@citea\def\@citea{,}\@ifundefined
	{b@\@citeb}{{\bf ?}\@warning
	{Citation `\@citeb' on page \thepage \space undefined}}
	{\csname b@\@citeb\endcsname}}}{#1}}

\newif\if@cghi
\def\cite{\@cghitrue\@ifnextchar [{\@tempswatrue
	\@citex}{\@tempswafalse\@citex[]}}
\def\citelow{\@cghifalse\@ifnextchar [{\@tempswatrue
	\@citex}{\@tempswafalse\@citex[]}}
\def\@cite#1#2{{$\null^{#1}$\if@tempswa\typeout
	{IJCGA warning: optional citation argument
	ignored: `#2'} \fi}}

\def\pmb#1{\setbox0=\hbox{#1}
	\kern-.025em\copy0\kern-\wd0
	\kern.05em\copy0\kern-\wd0
	\kern-.025em\raise.0433em\box0}

\def\fnt#1#2{\footnotetext{\kern-.3em
	{$^{\mbox{\scriptsize #1}}$}{#2}}}

\font\tenrm=cmr10
\font\tenit=cmti10
\font\tenbf=cmbx10
\font\bfit=cmbxti10 at 10pt
\font\ninerm=cmr9
\font\nineit=cmti9
\font\ninebf=cmbx9
\font\eightrm=cmr8

\def\qed{\hbox{${\vcenter{\vbox{			%HOLLOW SQUARE
   \hrule height 0.4pt\hbox{\vrule width 0.4pt height 6pt
   \kern5pt\vrule width 0.4pt}\hrule height 0.4pt}}}$}}

\newcommand{\asi}{\alpha^{\mbox{\tiny (1)}}_{\mbox{\scriptsize s}}}
\newcommand{\asit}{{\widetilde{\alpha}^{\mbox{\tiny
            (1)}}}_{\mbox{\scriptsize s}}}
\newcommand{\aan}{{\alpha}_{\mbox{\scriptsize an}}}
\newcommand{\aani}{\alpha^{\mbox{\tiny (1)}}_{\mbox{\scriptsize an}}}
\newcommand{\aanit}{{\widetilde{\alpha}^{\mbox{\tiny
            (1)}}}_{\mbox{\scriptsize an}}}
\newcommand{\myaan}{{^{\mbox{\tiny N}}}\!\aan}
\newcommand{\myaani}{{^{\mbox{\tiny N}}}\!\aani}
\newcommand{\myaanit}{{^{\mbox{\tiny N}}}\aanit}
\newcommand{\bst}{{\widetilde{\beta}}_{\mbox{\scriptsize s}}}
\newcommand{\myban}{{^{\mbox{\tiny N}}}\!{\beta}_{\mbox{\scriptsize an}}}
\newcommand{\mybant}{{^{\mbox{\tiny
            N}}}\!{\widetilde{\beta}}_{\mbox{\scriptsize an}}}
\begin{document}
\setlength{\textheight}{7.7truein}  %for 2nd page onwards

%\runninghead{}{}

\normalsize\textlineskip
\pagestyle{empty}
\setcounter{page}{1}

\copyrightheading{}			%{Vol. 0, No.0 (1992) 000--000}

\vspace*{0.88truein}

%\fpage{1}
\centerline{\bf INVESTIGATION OF A NEW ANALYTIC}
\baselineskip=13pt
\centerline{\bf RUNNING COUPLING IN QCD}
\vspace*{0.37truein}
\centerline{\footnotesize A.\ V.\ NESTERENKO\footnote{
Electronic address: nesterav@thsun1.jinr.ru}}
\baselineskip=12pt
\centerline{\footnotesize\it Physical Department, Moscow State University}
\baselineskip=10pt
\centerline{\footnotesize\it Moscow, 119899, Russia}
\vspace*{0.225truein}

%\publisher{(received date)}{(revised date)}

\vspace*{0.21truein}
\abstracts{
The mathematical properties of the new analytic running coupling (NARC) in
QCD are investigated. This running coupling naturally arises under
``analytization'' of the renormalization group equation.  One of the
crucial points in our consideration is the relation established between
the NARC and its inverse function. The latter is expressed in terms of the
so-called Lambert $W$ function. This relation enables one to present
explicitly the NARC in the renorminvariant form and to derive the
corresponding $\beta$ function. The asymptotic behavior of this $\beta$
function is examined. The consistent estimation of the parameter
$\Lambda_{\mbox{\tiny QCD}}$ is given.}{}{}

%\vspace*{10pt}
%\keywords{PACS number(s): 12.38.Lg, 02.30.Gp, 12.38.Aw}

\vspace*{1pt}\textlineskip
\section{Introduction}
     The description of hadron interactions at small characteristic
momenta remains an actual problem of elementary particle theory. The
asymptotic freedom in quantum chromodynamics (QCD) enables one to
investigate processes at large momentum transfers by making use of
standard perturbation theory. But there is a number of phenomena which
description lies beyond such calculations, namely quark confinement, gluon
and quark condensates and many others. For these purposes nonperturbative
approaches are used.

     In the late 50's the so-called analytic approach to quantum
electrodynamics was proposed.\cite{Redm,BLS} Its basic idea is the
explicit imposition of the causality condition which implies the
requirement of the analyticity in the $Q^2$ variable for the relevant
physical quantities. Recently this approach has been extended to
QCD.\cite{ShSol} This leads to the following essential advantages: absence
of unphysical singularities at any loop level, stability in the infrared
(IR) region, stability with respect to loop corrections, and extremely
weak scheme dependence. The analytic approach has been applied
successfully to such QCD problems as the $\tau$ lepton decays,
$e^+e^-$-annihilation into hadrons, sum rules.\cite{SolSh}

     Recently the analytic approach has been employed to the
renormalization group (RG) equation.\cite{PRD,Austria} The analyticity
requirement was imposed on the RG equation itself, before deriving its
solution.  Solving the RG equation, ``analytized'' in the above-mentioned
way (i.e., the proper analytical properties of the RG equation as a whole
have been recovered), one arrives at the new analytic running
coupling\cite{PRD} (NARC). An essential point, that plays a crucial role
in description of a number of nonperturbative phenomena in our approach,
is the IR enhancement of the new analytic running coupling at $Q^2=0$. It
should be stressed here that such a behavior of the invariant charge is in
agreement with the Schwinger--Dyson equations,\cite{AlekArbu} and, as it
was demonstrated recently,\cite{PRD,Austria} provides description of quark
confinement {\it without invoking any additional assumptions}.

     The objective of this paper is to study in details the mathematical
properties of the new analytic running coupling in QCD.\cite{PRD} This
enables one to present the NARC in the explicitly renorminvariant form and
to construct manifestly the corresponding $\beta$ function. We restrict
ourself to the one-loop level calculations, taking into account the
following consideration. Due to the higher loop stability\cite{PRD} all
the essential features of the new analytic running coupling become
apparent at the one-loop level already. The account of higher--loop
corrections does not lead to any qualitative changes. At the same time, at
the one-loop level one able to carry out {\it all the calculations in an
explicit form}, that is one of the self--evident merits of the current
consideration.

     The layout of the paper is as follows. In Sec.\ 2 the new analytic
running coupling is briefly discussed. In Sec.\ 3 the function $N(a)$,
that plays a key role in our analysis, is introduced and its properties
are investigated. In turn, the function $N(a)$ is expressed in terms of
the so-called Lambert $W$ function. Proceeding from this, in Sec.\ 4 the
one-loop new analytic running coupling is presented manifestly in the
renorminvariant form and its basic properties are revealed. Further, the
corresponding $\beta$ function is derived in an explicit form and its
properties are examined. The consistent estimation of the parameter
$\Lambda_{\mbox{\tiny QCD}}$ is given. In the Conclusion (Sec.~5) the
obtained results are formulated in a compact way, and further studies in
this approach are outlined.

\section{New analytic running coupling in QCD}
     In the analytic approach, extended to QCD by Shirkov and
Solovtsov,\cite{ShSol} the basic idea is the explicit imposing of the
causality condition, which implies the requirement of the analyticity in
the $Q^2$ variable for the relevant physical quantities. Later this idea
was applied to the ``analytization'' of the perturbative series (i.e., the
recovering of the proper analyticity properties of this series) when
calculating the QCD observables.\cite{SolSh}

     As known, the $Q^2$-evolution of QCD observables is usually described
in the framework of the RG approach. However, the standard RG calculations
involve, as a rule, the use of the perturbative expansions. As a result,
the RG equation at any given loop level does not meet the analyticity
requirement.\cite{PRD,Austria} In order to avoid such situation, the
procedure of analytization of the RG equation as a whole has been
elaborated.\cite{PRD,MyDipl} This procedure implies the recovering of the
correct analytic properties of the RG equation itself, before deriving its
solution. As a result, the solution of the analytized RG equation is
expressed in terms of a new analytic running coupling. At the one-loop
level this running coupling reads\cite{PRD}
\begin{equation}
\label{NARCDef}
\myaani(Q^2) =
\frac{4\pi}{\beta_0}\,\frac{z-1}{z\,\ln z},\quad
z = \frac{Q^2}{\Lambda^2},
\end{equation}
where $\beta_0=11-2\,n_{\mbox{\scriptsize f}}/3$ is the first coefficient
of the $\beta$ function, $n_{\mbox{\scriptsize f}}$ is the number of
active quarks. At the higher loop levels only the integral representation
for $\myaan(Q^2)$ has been obtained.\cite{PRD} So, at the $k$-loop level
we have
\begin{equation}
\label{NARCDefHL}
^{\mbox{\tiny N}}\!\alpha^{(k)}_{\mbox{\scriptsize an}}(Q^2) =
{^{\mbox{\tiny N}}}\!\alpha^{(k)}_{\mbox{\scriptsize an}}(Q^2_0)
\,\frac{z_0}{z}\,
\exp\!\left[\int_{0}^{\infty}
{\cal R}^{(k)}(\sigma)
\,\ln\!\left(\frac{\sigma+z}{\sigma+z_0}\right)
\frac{d\sigma}{\sigma}\right],
\end{equation}
where ${\cal R}^{(k)}(\sigma) = (2 \pi i)^{-1}
\sum_{j=0}^{k-1}\frac{\beta_{j}}{\beta_{0}^{j+1}} \left\{\left[
\widetilde{\alpha}^{(k)}_{\mbox{\scriptsize s}}(-\sigma-i\varepsilon)
\right]^{j+1}-\left[
\widetilde{\alpha}^{(k)}_{\mbox{\scriptsize s}}(-\sigma+i\varepsilon)
\right]^{j+1}\right\}$ is the spectral density,
$\widetilde{\alpha}^{(k)}_{\mbox{\scriptsize s}}$ is the $k$-loop
perturbative running coupling, $\widetilde{\alpha} \equiv \alpha\,
\beta_0/(4\pi)$, and $z_0=Q^2_0/\Lambda^2$ is the normalization point.

\begin{figure}[ht] %ORIGINAL SIZE: width=1.4TRUEIN; height=1.5TRUEIN
%\psdraft
\noindent
\vspace*{13pt}
\centerline{\epsfig{file=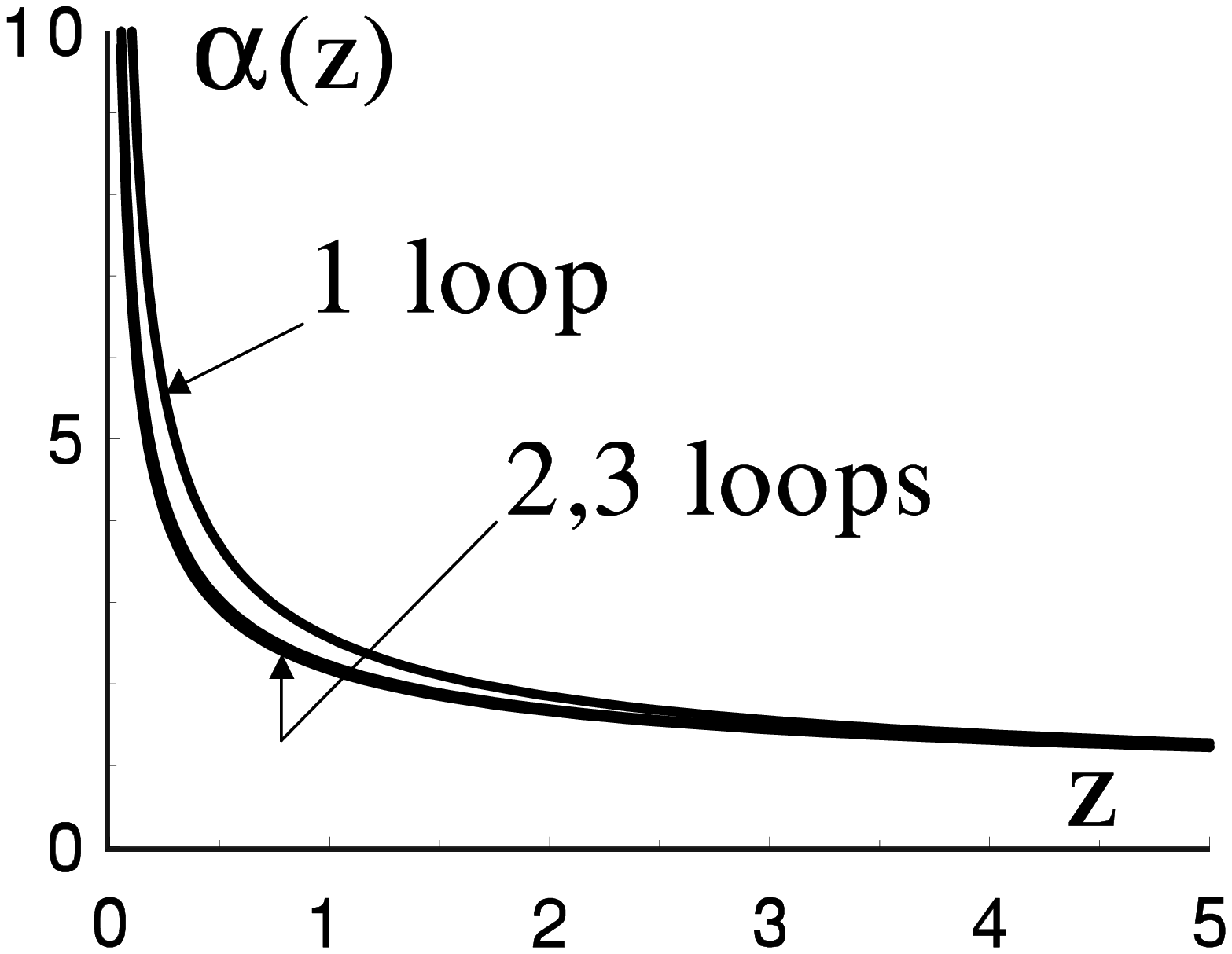, width=90mm}}
\vspace*{13pt}
\fcaption{The normalized new analytic running coupling $\alpha(z) =
{^{\mbox{\tiny N}}}\!{\alpha}_{\mbox{\scriptsize an}}(Q^2)\,
/\,{^{\mbox{\tiny N}}}\!{\alpha}_{\mbox{\scriptsize
an}}(Q^2_0)$ at the one-, two-, and three-loop levels. The
normalization point is $Q^2_0=10\,\Lambda^2$, $\,z=Q^2/\Lambda^2$.}
\label{NARCPlot}
\end{figure}

     It is worth making here a short historical comment. When considering
the analytic approach to quantum electrodynamics, it was noted that the
method proposed in Refs.\cite{Redm,BLS} is not unique. For example, in
paper\cite{Kirzh} the ambiguity of that method was shown explicitly.
Furthermore, it was stressed that for avoiding the ambiguity of the
approach developed in Refs.\cite{Redm,BLS} one should impose an additional
condition,\cite{Arbu} in particular, by making use of the equations of
motion. Specifically, following this way the boson propagator has been
derived\cite{Arbu} in which the singularity was removed in a
multiplicative way (i.e., similar to Eq.~(\ref{NARCDef})). Moreover,
another common feature of our consideration and that of paper\cite{Arbu}
is the following. Unlike the method,\cite{BLS,ShSol} in our approach one
is also able to catch the nonperturbative contributions to the spectral
density for the running coupling (see also Eq.~(\ref{NARCIntRep})).

     Figure~\ref{NARCPlot} presents the new analytic running coupling
calculated at the one-, two- and three-loop levels. These curves show
that the analytic running coupling~(\ref{NARCDef}) possesses the higher
loop stability. Moreover, it can be demonstrated that the singularity of
the NARC at the point $Q^2=0$ is of the universal type at any loop
level.\cite{PRD}

\section{The Lambert $W$ function}
\label{LambN}
     For the representation of the new analytic running coupling
(\ref{NARCDef}) in the renorminvariant form and for the construction of
the relevant $\beta$ function, it proves to be convenient to use the
so-called Lambert $W$ function. As long ago as the middle of $18$th
century this function is being employed in diverse physical
problems.\cite{Corless} In current researches in QCD the interest to this
function arose just a few years ago. So, it was revealed\cite{Grunb,Magr}
that the exact solution to the perturbative RG equation for the invariant
charge at the two-loop approximation can be expressed in terms of the
Lambert $W$ function. At the three-loop level, when the Pad\'e
approximation for the $\beta$ function is used, the solution to this
equation can be expressed in terms of the Lambert $W$ function
also.\cite{Grunb} But in this case one has to take into account that the
application of the Pad\'e approximation to the $\beta$ function
drastically changes the behavior of perturbative running coupling at small
$Q^2$.

     The Lambert $W$ function is defined as a many--valued function
$W(x)$, that satisfies the equation
\begin{equation}
\label{WEqDef}
W(x)\,\exp\left[W(x)\right]=x.
\end{equation}
\begin{figure}[ht] %ORIGINAL SIZE: width=1.4TRUEIN; height=1.5TRUEIN
%\psdraft
\noindent
\vspace*{13pt}
\centerline{\epsfig{file=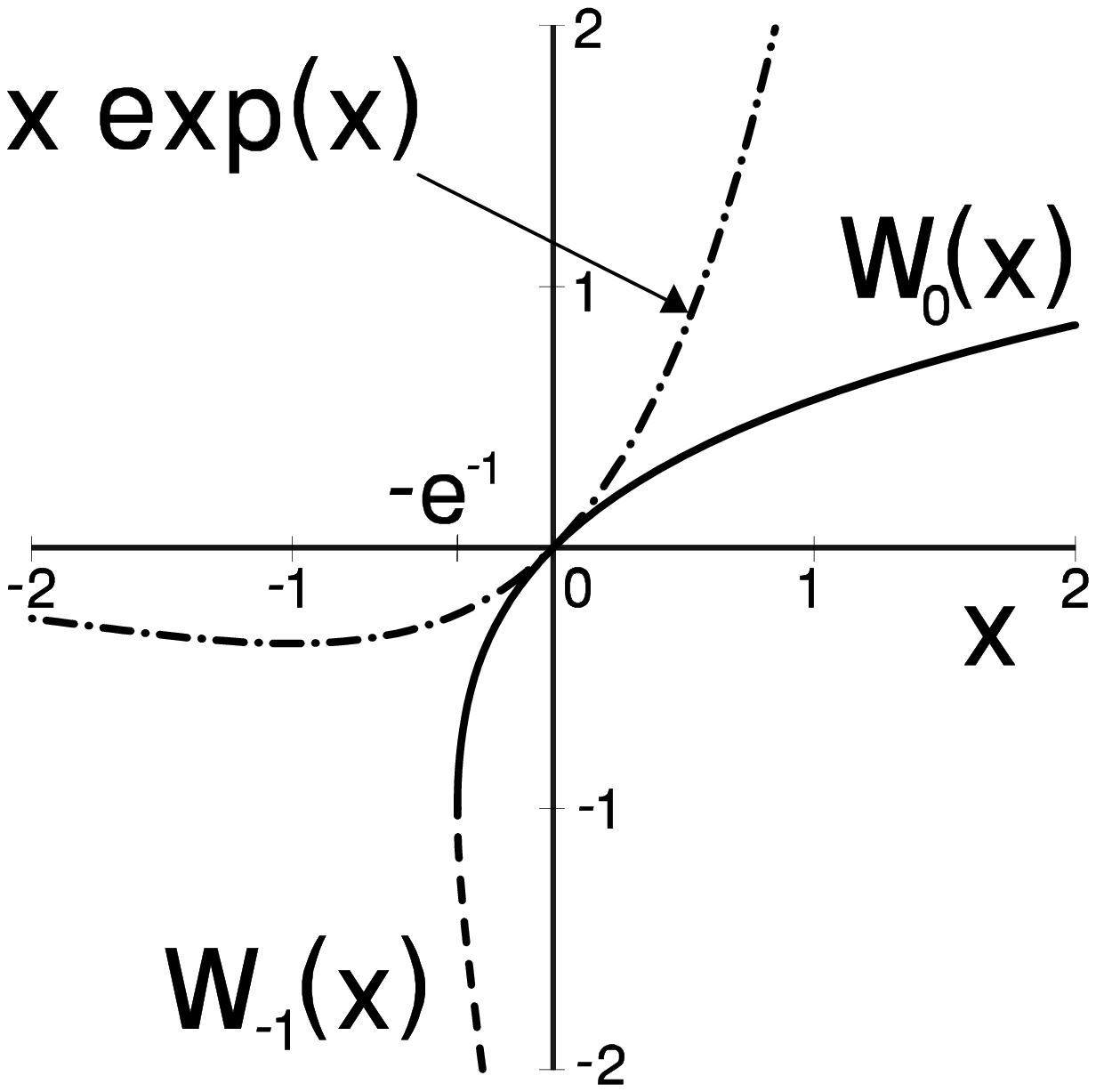, width=65mm}}
\vspace*{13pt}
\fcaption{The function $x\,e^x$ (dot-dashed curve) and two real branches,
$W_{0}(x)$ (solid curve) and $W_{-1}(x)$ (dashed curve) of its inverse
function.}
\label{LambertPlot}
\end{figure}

\noindent
Only two real branches of this function, the principal branch $W_0(x)$ and
the branch $W_{-1}(x)$ (see Fig.~\ref{LambertPlot}), will be used in our
consideration. The other branches of the Lambert $W$ function take an
imaginary values. One can show that for the branches $W_0$ and $W_{-1}$
the following expansions hold:
\begin{eqnarray}
\label{WZSerOr}
W_{0}(\varepsilon) &=& \varepsilon - \varepsilon^2 + O(\varepsilon^3),
\quad \varepsilon \to 0, \\
\label{WUSerOr}
W_{-1}(-\varepsilon) &=& \ln \varepsilon + O(\ln |\ln\varepsilon|),
\quad \varepsilon \to 0_{+}, \\
\label{WZSerBr}
W_{0}\left(-\frac{1}{e} + \varepsilon \right) &=& -1 + \sqrt{2 e
\varepsilon} + O(\varepsilon), \quad \varepsilon \to 0_{+},\\
\label{WUSerBr}
W_{-1}\left(-\frac{1}{e} + \varepsilon \right) &=& -1 - \sqrt{2 e
\varepsilon} + O(\varepsilon), \quad \varepsilon \to 0_{+}.
\end{eqnarray}
Details concerning the mathematical properties of the Lambert $W$ function
can be found in the review.\cite{Corless}

\newpage
     In order to represent the NARC~(\ref{NARCDef}) in a renorminvariant
form and to derive the $\beta$ function corresponding to $\myaani$ one has
to solve the equation (see Sec.~4)
\begin{equation}
\label{EqGen}
\frac{z-1}{z\,\ln z} = a,\quad z>0,\quad a>0
\end{equation}
with respect to the variable $z$.
Let us multiply this equation through by the factor $a^{-1}\,\ln z$,
\begin{equation}
\label{EqMod}
\ln z = \frac{b}{z} - b,\quad z>0,\quad b = -\frac{1}{a}<0.
\end{equation}
The equation obtained has a trivial solution $z=1$ that does not satisfy
initial Eq.~(\ref{EqGen}) when $a \neq 1$. Obviously, this is a
consequence of the multiplication of Eq.~(\ref{EqGen}) by $\ln z$.
Therefore, when solving Eq.~(\ref{EqMod}) for the variable $z$ one has to
discard its trivial solution $z=1$. Next, we can represent
Eq.~(\ref{EqMod}) in the form
\begin{equation}
\label{EqModMod}
\frac{b}{z}\,\exp\!\left(\frac{b}{z}\right) = b\,e^{b}.
\end{equation}
Taking into account the definition~(\ref{WEqDef}), solution to Eq.\
(\ref{EqModMod}) can be expressed in terms of the Lambert $W$ function
\begin{equation}
\label{EqModSol}
\frac{b}{z} = W_k \left(b\,e^b\right).
\end{equation}
The branch index $k$ of the $W$ function will be specified below.

     In the physically relevant range $a>0$ the argument of the Lambert
$W$ function in Eq.~(\ref{EqModSol}) takes the values $-1/e\le b\,e^{b} <
0$. The only two real branches, the principle branch $W_0$ and the branch
$W_{-1}$ (see Fig.~\ref{LambertPlot}) correspond to this interval. Here
one has to handle carefully with the interchange between branches $W_0(x)$
and $W_{-1}(x)$ at the point $x=-1/e$ (this corresponds to the value
$a=1$).

\begin{figure}[ht] %ORIGINAL SIZE: width=1.4TRUEIN; height=1.5TRUEIN
%\psdraft
\noindent
\vspace*{13pt}
\centerline{\epsfig{file=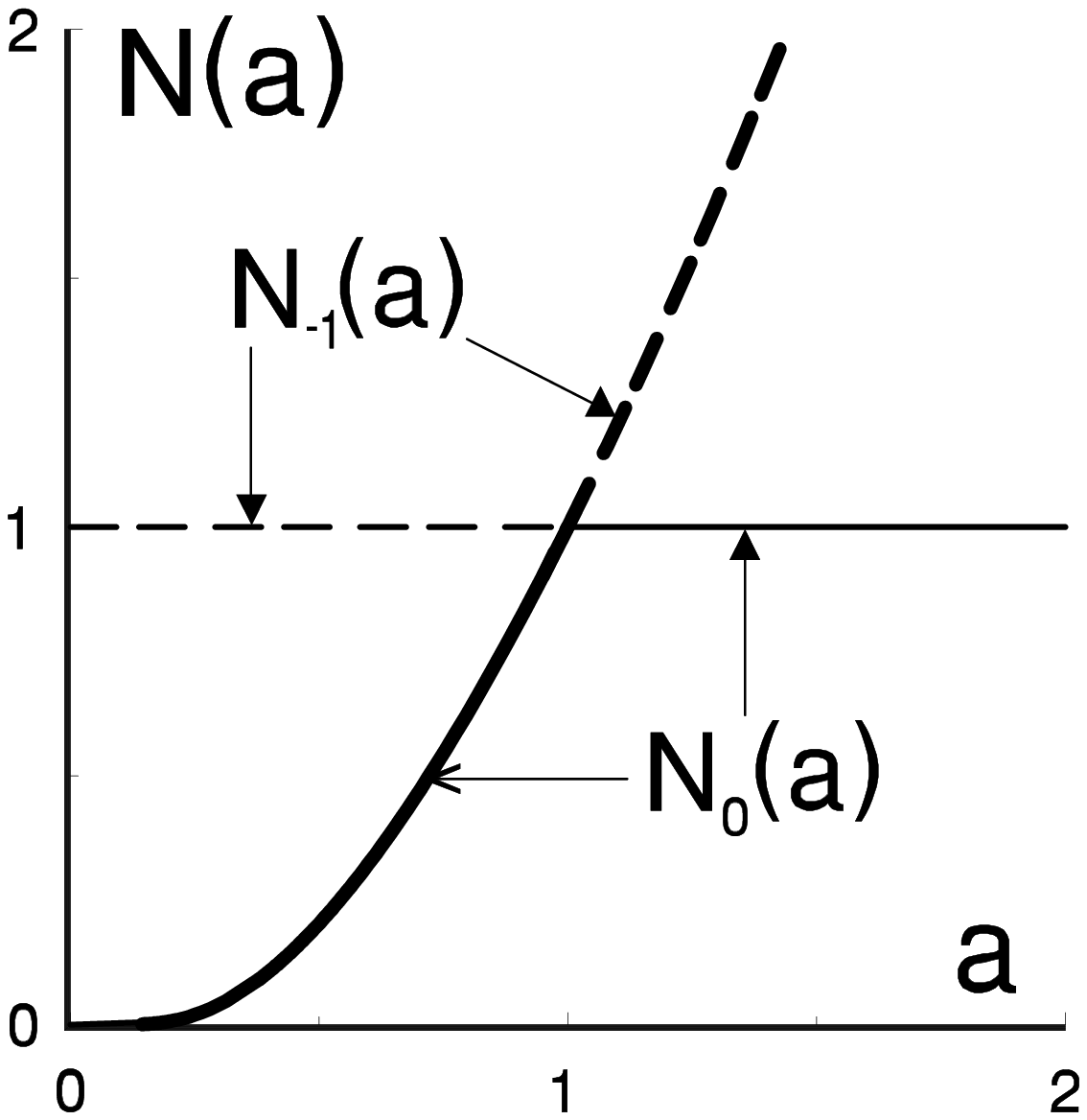, width=65mm}}
\vspace*{13pt}
\fcaption{The function $N(a)$ (bold curve), and the functions $N_{0}(a)$
(solid curve) and $N_{-1}(a)$ (dashed curve) (see
Eq.~(\protect\ref{NDef})).}
\label{NPlot}
\end{figure}

     So, the nontrivial solution to Eq.~(\ref{EqMod}) for the
variable $z$ is
\begin{equation}
\label{EqModSolW}
\frac{1}{z} = \cases{
b^{-1}\, W_{0}\!\left(b\,e^b\right), & $b < -1$, \cr
b^{-1}\, W_{-1}\!\left(b\,e^b\right), & $-1 \le b < 0$ \cr}
\end{equation}
(another choice of branches will be considered below). Thus, the
solution to Eq.~(\ref{EqGen}) we are interested in can be written
in the form:
\begin{equation}
\label{EqGenSol}
z = \frac{1}{N(a)},
\end{equation}
where we have introduced the function $N(a)$ (see Fig.~\ref{NPlot})
\begin{equation}
\label{NDef}
N(a) = \cases{
N_{0}(a),  & $0 < a \le 1$, \cr
N_{-1}(a), & $a > 1$, \cr}
\quad
N_{k}(a) = -a\, W_{k}\!
\left[-\frac{1}{a}\,\exp\!\left(-\frac{1}{a}\right)\right].
\end{equation}

     As follows from the definition~(\ref{WEqDef}) the function $xe^x$ and
the Lambert $W$ function are mutually inverse. Here one has to distinguish
precisely the branches $W_0$ and $W_{-1}$ (see Fig.~\ref{LambertPlot}).
From this figure it follows directly that for $x > -1$ the function $xe^x$
``corresponds'' to the branch $W_0$, and for $x \le -1$ the function
$xe^x$ ``corresponds'' to the branch $W_{-1}$. Let us introduce for
convenience the notations of the inverse functions for these branches
\begin{equation}
\label{ArcWDef}
x\,e^x = \cases{
\left(W_{0}\right)^{-1}\!(x),  & $x > -1$, \cr
\left(W_{-1}\right)^{-1}\!(x), & $x \le -1$. \cr}
\end{equation}
Now the function $N(a)$ in Eq.~(\ref{NDef}) can be written in a more
compact form
\begin{equation}
\label{NDefArcW}
N(a) = \cases{
-a\,W_{0}\!\left[\left(W_{-1}\right)^{-1}\!\left(-a^{-1}\right)\right], &
$0 < a \le 1$, \cr
-a\,W_{-1}\!\left[\left(W_{0}\right)^{-1}\!\left(-a^{-1}\right)\right], &
$a > 1$. \cr}
\end{equation}
It is worth noting that another choice of the branches of the function
$N(a)$
\begin{equation}
\label{NPrDef}
\widetilde{N}(a) = \cases{
N_{-1}(a), & $0 < a \le 1$, \cr
N_{0}(a),  & $a > 1$ \cr}
\end{equation}
leads to the trivial solution of Eq.~(\ref{EqMod}) (in this case
$\widetilde{N}(a) \equiv 1$). Therefore, the solution
$z=1/\widetilde{N}(a)$ does not satisfy Eq.~(\ref{EqGen}) when $a\neq 1$,
and consequently we have to reject it.

     Let us briefly consider the basic properties of the function $N(a)$
introduced in Eq.~(\ref{NDef}). In the physical range $a>0$ it is a
nonnegative, monotone increasing function (see Fig.~\ref{NPlot}). With the
help of the series~(\ref{WZSerOr})--(\ref{WUSerBr}) one can show that for
the function $N(a)$ the following expansions hold
\begin{eqnarray}
\label{NSerOr}
N(a\to 0_{+}) &=& \exp\!\left(-\frac{1}{a}\right)
\left\{1 + O\!\left[\exp\!\left(-\frac{1}{a}\right)\right]\right\}; \\
\label{NSerUn}
N(1+\varepsilon) &=& 1 + 2\varepsilon + O(\varepsilon^2), \quad
\varepsilon \to 0; \\
\label{NSerInf}
N(a\to\infty) &=& a\ln a + O\left[a\ln(\ln a)\right].
\end{eqnarray}

\section{Properties of the new analytic running coupling}
\label{NARCProperties}
     In this section we turn to the investigation of the properties of
the one-loop new analytic running coupling in QCD\cite{PRD}
\begin{equation}
\label{NARCDefNew}
\myaani (Q^2) = \frac{4\pi}{\beta_0}\,\frac{z-1}{z\,\ln z}, \quad
z=\frac{Q^2}{\Lambda^2}.
\end{equation}
For the convenience we omit the group factor and proceed with the
expression
\begin{equation}
\label{ADef}
a(Q^2) \equiv \myaanit (Q^2) = \frac{z-1}{z\,\ln z}.
\end{equation}
Let us consider first of all the asymptotics of this function. In the
ultraviolet (UV) limit the standard behavior of the invariant charge is
reproduced: $a(Q^2~\!\to~\!\infty)\to1/\ln z$, $\,z=Q^2/\Lambda^2$, i.e., the
asymptotic freedom of the theory is taken into account. In the IR region
there is an enhancement of the running coupling~(\ref{ADef}): $a(Q^2 \to
0_{+})\to -[\varepsilon \ln\varepsilon]^{-1}$, $\,\varepsilon =
Q^2/\Lambda^2$. One should note here that such a behavior of the invariant
charge is in agreement with the Schwinger--Dyson equations,\cite{AlekArbu}
and as it was demonstrated recently,\cite{PRD,Austria} provides quark
confinement {\it without invoking any additional assumptions}. Thus the
running coupling~(\ref{NARCDefNew}) incorporates both the asymptotic
freedom behavior and the IR enhancement in a single expression. Obviously
it is an essential advantage of our approach. It is easy to show that the
function~(\ref{ADef}) has smooth behavior in the vicinity of the point
$Q^2=\Lambda^2$: $a(Q^2~\!\to~\!\Lambda^2) = 1-\delta/2+O(\delta^2)$,
$\,\delta=Q^2/\Lambda^2 - 1$. It is worth noting also that for the running
coupling~(\ref{ADef}) the causal representation of the K\"all\'en-Lehmann
type holds\cite{PRD}
\begin{equation}
\label{NARCIntRep}
\myaanit(Q^2)=\int_{0}^{\infty}
\frac{^{\mbox{\tiny N}}\!\rho(\sigma)}{\sigma+z}\,d\sigma,\quad
^{\mbox{\tiny N}}\!\rho(\sigma)=
\left(1+\frac{1}{\sigma}\right)\frac{1}{\ln^2\sigma+\pi^2}.
\end{equation}

     In general, any expression for the running coupling makes sense only
if the relevant definition of the parameter $\Lambda$ is provided.
Otherwise, a running coupling may be not a renorminvariant quantity at
all. Thus, it is of a primary importance to present both the running
coupling and the corresponding definition for its parameter $\Lambda$
explicitly.

     So, let us represent the running coupling $\myaani$ in the
renorminvariant form. In the general case for the invariant charge $\bar
g^2(Q^2/\mu^2, g)/(4\pi) \equiv \alpha(Q^2)$ the normalization condition
$\bar g^2(1, g) = g^2$ must be fulfilled.\cite{Ynd} In our case this
normalization condition acquires the form
\begin{equation}
\label{NormCon}
\frac{(\mu^2/\Lambda^2)-1}{\ln(\mu^2/\Lambda^2)\,\mu^2/\Lambda^2} =
\frac{\beta_0}{4 \pi} \alpha(\mu^2) \equiv a(\mu^2).
\end{equation}
Therefore, we have to resolve this equation with respect to the parameter
$\Lambda$. The solution to such equations was considered in details in
Sec.~3. With the help of the function $N(a)$ (see
Eq.~(\ref{NDef})) the solution to Eq.~(\ref{NormCon}) can be presented in
the form
\begin{equation}
\label{LRGInv}
\Lambda^2 = \mu^2\, N\!\left[\frac{\beta_0}{4 \pi} \alpha(\mu^2)\right].
\end{equation}
Thus, the renorminvariant expression for the one-loop new analytic
running coupling~(\ref{NARCDef}) is the following
\begin{equation}
\label{NARCRGInv}
\myaani(Q^2) = \frac{4 \pi}{\beta_0}\,\frac{z-1}{z\,\ln z}, \quad
z = \frac{Q^2}{\Lambda^2}, \quad
\Lambda^2 = \mu^2\, N\!\left[\frac{\beta_0}{4 \pi} \alpha(\mu^2)\right].
\end{equation}
It is worth noting that when $\mu\to\infty$ (see Eq.~(\ref{NSerOr})) the
right hand side of Eq.~(\ref{LRGInv}) tends to its standard form\cite{Ynd}
corresponding to the perturbative running coupling
$\asi(Q^2)=4\pi/(\beta_0\ln z)$
\begin{equation}
\label{LRGInvPert}
\Lambda_{\mbox{\scriptsize s}}^2 = \mu^2\, \exp\!\left[-\frac{4 \pi}{\beta_0}
\frac{1}{\alpha(\mu^2)}\right].
\end{equation}

     Let us proceed to the construction of the $\beta$ function
corresponding to the running coupling~(\ref{NARCDef}). By
definition,\cite{QCD}
\begin{equation}
\label{BetaGenDef}
\beta(\alpha) = \frac{\partial\,\alpha(\mu^2)}{\partial\,\ln\mu^2}.
\end{equation}
In the case under consideration, for the invariant charge $\myaanit (Q^2)$
the $\beta$ function takes the form
\begin{equation}
\label{NABetaTent}
\mybant(a) = \left[\frac{1}{z(a)} - a\right]\frac{1}{\ln [z(a)]}\, ,
\end{equation}
where $z(a)$ is the solution of Eq.~(\ref{EqGen}) with respect to the
variable $z$ (see Sec.~3). For the convenience we omit here the
group factor, and use the expression $\mybant(a) = \myban (a)\,
\beta_0/(4\pi)$. With the help of the function $N(a)$ defined in
Eq.~(\ref{NDef}) the $\beta$ function for the NARC~(\ref{ADef}) can be
written in the form
\begin{equation}
\label{NABetaDef}
\mybant(a) = \frac{a - N(a)}{\ln \left[N(a)\right]}\, .
\end{equation}
\begin{figure}[ht] %ORIGINAL SIZE: width=1.4TRUEIN; height=1.5TRUEIN
%\psdraft
\noindent
\vspace*{13pt}
\centerline{\epsfig{file=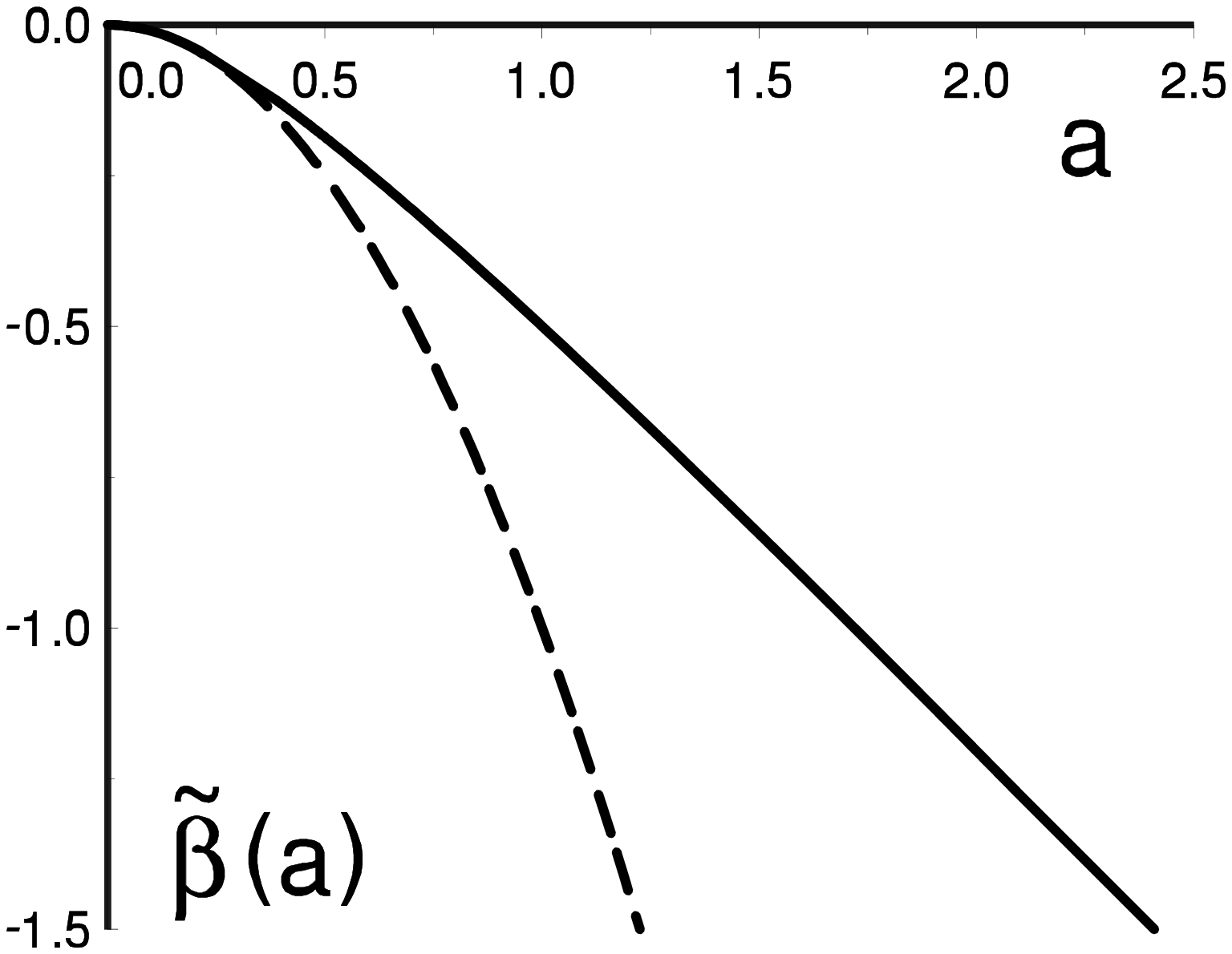, width=85mm}}
\vspace*{13pt}
\fcaption{The solid curve presents the $\beta$ function $^{\mbox{\tiny
N}}{\widetilde{\beta}}_{\mbox{\scriptsize an}}(a)$ (Eq.\
(\protect\ref{NABetaDef})) corresponding to the one-loop new analytic
running coupling $^{\mbox{\tiny N}}{\widetilde{\alpha}^{\mbox{\tiny
(1)}}}_{\mbox{\scriptsize an}}$ (Eq.\ (\protect\ref{ADef})). The standard
perturbative result for this function
${\widetilde{\beta}}_{\mbox{\scriptsize s}}(a)=-a^2$ is also shown (dashed
curve).}
\label{BetaPlot}
\end{figure}

\noindent
Figure~\ref{BetaPlot} presents this $\beta$ function and the standard
perturbative result $\bst (a) = - a^2$ corresponding to the one-loop
perturbative running coupling $\asit(Q^2) = 1/\ln z$. Thus, the $\beta$
function corresponding to the one-loop new analytic running coupling
$\myaani$ reads as
\begin{equation}
\label{NABetaDefGen}
\myaani(Q^2) = \frac{4\pi}{\beta_0}\,\frac{z-1}{z\,\ln z} \equiv
\frac{4\pi}{\beta_0}\,a(Q^2), \quad
\myban(a) = \frac{4\pi}{\beta_0}\,\frac{a - N(a)}{\ln \left[N(a)\right]}\, .
\end{equation}

     The numerator and denominator on the right hand side of
Eq.~(\ref{NABetaDef}) have the opposite signs for all positive~$a$. This
follows directly from the expansion~(\ref{NABetaSerUn}) (see below) and
from Figs.~\ref{NPlot} and~\ref{BetaPlot}. Thus, the $\beta$
function~(\ref{NABetaDef}) possesses an important property, namely
$\mybant(a) \le 0$ for all positive~$a$. It is this property that provides
the asymptotic freedom of the theory. By making use of the series
(\ref{NSerOr})--(\ref{NSerInf}) one can derive the following expansions of
the $\beta$ function~(\ref{NABetaDef}). First of all,
\begin{equation}
\label{NABetaSerOr}
\mybant(a \to 0_{+}) = -a^2 +
O\!\left[\exp\!\left(-\frac{1}{a}\right)\right],
\end{equation}
i.e., the standard perturbative limit is reproduced in the UV region.
Secondly,
\begin{equation}
\label{NABetaSerUn}
\mybant(1+\varepsilon) = - \frac{1}{2} - \frac{2}{3}\varepsilon +
O(\varepsilon^2),\quad \varepsilon \to 0,
\end{equation}
i.e., the function~(\ref{NABetaDef}) has a smooth behavior in the vicinity
of the point $a=1$. Next, in the IR region
\begin{equation}
\label{NABetaSerInf}
\mybant(a\to\infty) = -a \left[1 +
O\!\left(\frac{\ln(\ln a)}{\ln^2 a}\right)\right].
\end{equation}
Such a behavior of the $\beta$ function provides the IR enhancement of the
new analytic running coupling. This enhancement leads ultimately to the
quark--antiquark potential rising at large distances.\cite{PRD,Austria}

     Thus, as one could anticipate, the asymptotics of the obtained
$\beta$ function incorporate both the asymptotic freedom and the IR
enhancement of the running coupling~(\ref{NARCDef}).

     Sometimes it is more convenient to present the $\beta$ function in
terms of the invariant charge $g$, $\alpha \equiv g^2/(4\pi)$. In this
case the asymptotic expansions~(\ref{NABetaSerOr}) and
(\ref{NABetaSerInf}) can be rewritten in the following way: $\myban (g)
\simeq - \beta_0\,g^3/(16 \pi^2)$ when $g \to 0$ (this is nothing but the
well--known standard perturbative result\cite{QCD}); and $\myban (g)
\simeq -g$ when $g \to \infty$. It is worth noting here that a similar
(i.e., linear in $g$) asymptotic behavior $\beta (g) \simeq - 2 g$ when $g
\to \infty$ holds, when the so-called Mandelstam approximation is used
(see paper\cite{Smekal} and references therein). However the latter leads
to the running coupling with more singular behavior in the IR region,
namely $\alpha (Q^2) \sim Q^{-4}$, when $Q^2 \to 0_{+}$.

     As it was mentioned in the previous section, the new analytic running
coupling~(\ref{NARCDef}) incorporates both the asymptotic freedom and the
IR enhancement in a single expression. Obviously this is an essential
advantage of our approach. It was shown explicitly\cite{PRD,Austria} that
invariant charge of the form~(\ref{NARCDef}) leads to the rising
quark--antiquark ($q\bar q$) potential without invoking any additional
assumptions. The comparison of this $q\bar q$ potential with the so-called
Cornell phenomenological potential gives the following estimation for the
$\Lambda_{\mbox{\tiny QCD}}$ parameter: $\Lambda = (0.60 \pm 0.1)$~GeV.
The fit has been performed in the physically meaning
region\cite{Brambilla} $0.1 \: \mbox{fm} \le r \le 1.0 \: \mbox{fm}$ (in
this interval three active quarks should be taken into account,
$n_{\mbox{\scriptsize f}}=3$). The analogous comparison of the $q\bar q$
potential derived in Ref.\cite{PRD} with the lattice data\cite{Bali} has
been carried out recently. This gives a close value, $\Lambda = (0.57 \pm
0.1)$~GeV. Furthermore, the estimation of the parameter
$\Lambda_{\mbox{\tiny QCD}}$ has also been performed recently when
deducing the value of the gluon condensate by making use of the
NARC~(\ref{NARCDef}).  This gives the value $\Lambda = (0.65 \pm
0.05)$~GeV. Therefore, we have the acceptable estimation of the parameter
$\Lambda_{\mbox{\tiny QCD}}$ in the framework of our approach: $\Lambda =
(0.60 \pm 0.09)$~GeV.

     All this testifies that the new analytic running coupling
(\ref{NARCDef}) substantially involves the nonperturbative behavior of
quantum chromodynamics.

\section{Conclusion}
     The mathematical properties of the new analytic running coupling in
QCD~(\ref{NARCDef}) are investigated. This running coupling naturally
arises under analytization of the renormalization group
equation.\cite{PRD} One of the crucial points in our consideration is the
explicit relation established between the NARC (Eq.~(\ref{NARCDef})) and
its inverse function $N(a)$ (Eq.~(\ref{NDef})). The latter can be
expressed in terms of the so-called Lambert $W$ function. The properties
of the function $N(a)$ are examined in details. As known, when using the
RG approach it is important to present the relevant quantities in a
renorminvariant form. The function $N(a)$ enables one to perform this
explicitly for the NARC (\ref{NARCDef}). Furthermore, the $\beta$ function
corresponding to the running coupling $\myaani$ is expressed in terms of
the function $N(a)$. The explicit form of the function $\myban$ in
Eq.~(\ref{NABetaDefGen}) provides an essential advantage when considering
the gluon condensate. The asymptotics of the $\beta$ function in
Eq.~(\ref{NABetaDefGen}) are examined also. As one could anticipate, it
incorporates both the asymptotic freedom and the IR enhancement of the
invariant charge~(\ref{NARCDef}) in a single expression. It was shown that
there is a consistent estimation of the parameter $\Lambda_{\mbox{\tiny
QCD}}$ in the framework of our approach:  $\Lambda = (0.60 \pm 0.09)$~GeV.
Thus, the new analytic running coupling (\ref{NARCDef}) substantially
incorporates the nonperturbative behavior of the quantum chromodynamics.

     In further studies it would be interesting to investigate numerically
higher--loop corrections to the NARC~(\ref{NARCDef}) and to compare the
results with the explicit expressions obtained in the present paper.

\nonumsection{Acknowledgments}
     The author would like to thank Professor D.\ V.\ Shirkov and Dr.\ I.\
L.\ Solovtsov for interest in this work. The detailed discussions with
Dr.\ L.\ von Smekal are greatly appreciated. The author is grateful to
Professor G.\ S.\ Bali for supplying the relevant data on the lattice
calculations. The useful references provided by Professor R.\ M.\ Corless
and stimulating comments by Professor B.\ A.\ Arbuzov are thankfully
acknowledged.

     The partial support of RFBR (Grant No.\ 99-01-00091) is appreciated.

\nonumsection{References}

\end{document}